\documentclass[pdflatex,sn-mathphys-num]{sn-jnl}

\usepackage{graphicx}%
\usepackage{multirow}%
\usepackage{amsmath,amssymb,amsfonts}%
\usepackage{amsthm}%
\usepackage{mathrsfs}%
\usepackage[title]{appendix}%
\usepackage{xcolor}%
\usepackage{textcomp}%
\usepackage{manyfoot}%
\usepackage{booktabs}%
\usepackage{algorithm}%
\usepackage{algorithmicx}%
\usepackage{algpseudocode}%
\usepackage{listings}%
\usepackage{upgreek}
\usepackage{setspace}

\theoremstyle{thmstyleone}%
\theoremstyle{thmstyletwo}%

\theoremstyle{thmstylethree}%

\begin{document}

\title[Ultrafast giant enhancement of second harmonic generation in a strongly correlated cobaltite]{Ultrafast giant enhancement of second harmonic generation in a strongly correlated cobaltite}

\bgroup
\renewcommand{\email}[1]{} 
\author[1,2,3]{\fnm{Yuchen} \sur{Cui}}\email{cuiyuchen23@mails.ucas.ac.cn}
\equalcont{These authors contributed equally to this work.}
\author[4,5]{\fnm{Qiaomei} \sur{Liu}}\email{qiaomeiliu@buaa.edu.cn}
\equalcont{These authors contributed equally to this work.}

\author[4,6]{\fnm{Qiong} \sur{Wu}}\email{qwu@cqu.edu.cn}
\author[4]{\fnm{Shuxiang} \sur{Xu}}\email{shuxiangxu@pku.edu.cn}
\author[4]{\fnm{Junhan} \sur{Huang}}\email{jhhuang@stu.pku.edu.cn}
\author[4]{\fnm{Hao} \sur{Wang}}\email{hwang@stu.pku.edu.cn}
\author[4]{\fnm{Rongsheng} \sur{Li}}\email{lirsh@pku.edu.cn}
\author[7]{\fnm{Shanshan} \sur{Han}}\email{sshan@mail.nankai.edu.cn}

\author[1,2,3]{\fnm{Wei} \sur{Xu}}\email{xuwei@baqis.ac.cn}
\author[1,2,3]{\fnm{Li} \sur{Du}}\email{duli@baqis.ac.cn}

\author[1]{\fnm{Chunmei} \sur{Zhang}}\email{zhangcm@baqis.ac.cn}
\author[1]{\fnm{Ming} \sur{Lu}}\email{luming@baqis.ac.cn}
\author[1]{\fnm{Shangfei} \sur{Wu}}\email{wusf@baqis.ac.cn}
\author[2]{\fnm{Xinbo} \sur{Wang}}\email{xinbowang@iphy.ac.cn}
\author[4]{\fnm{Tao} \sur{Dong}}\email{taodong@sjtu.edu.cn}
\egroup
\author*[1]{\fnm{Li} \sur{Yue}}\email{yueli@baqis.ac.cn}
\author*[1,4,8,9]{\fnm{Nanlin} \sur{Wang}}\email{nlwang@sjtu.edu.cn}
\author*[1]{\fnm{Dong} \sur{Wu}}\email{wudong@baqis.ac.cn}

\affil*[1]{ \orgname{Beijing Academy of Quantum Information Sciences}, \orgaddress{\city{Beijing}, \postcode{100193}, \state{Beijing}, \country{China}}}

\affil[2]{\orgdiv{Beijing National Laboratory for Condensed Matter Physics}, \orgname{Institute of Physics, Chinese Academy of Sciences}, \orgaddress{, \city{Beijing}, \postcode{100190}, \state{Beijing}, \country{China}}}

\affil[3]{\orgname{University of Chinese Academy of Sciences}, \orgaddress{\city{Beijing}, \postcode{100049}, \state{Beijing}, \country{China}}}

\affil[4]{\orgdiv{International Center for Quantum Materials, School of Physics}, \orgname{Peking University}, \orgaddress{\city{Beijing}, \postcode{100871}, \state{Beijing}, \country{China}}}

\affil[5]{\orgdiv{School of Electronic and Information Engineering}, \orgname{Beihang University}, \orgaddress{\city{Beijing}, \postcode{100191}, \state{Beijing}, \country{China}}}

\affil[6]{\orgdiv{School of Physics and Center of Quantum Materials and Devices}, \orgname{Chongqing University}, \orgaddress{\city{Chongqing}, \postcode{401331}, \state{Chongqing}, \country{China}}}

\affil[7]{\orgdiv{School of Materials Science and Engineering}, \orgname{Nankai University}, \orgaddress{\city{Tianjin}, \postcode{300350}, \state{Tianjin}, \country{China}}}

\affil[8]{\orgname{Tsung-Dao Lee Institute, School of Physics and Astronomy, Shanghai Jiao Tong University}, \orgaddress{\city{Shanghai}, \postcode{201210}, \state{Shanghai}, \country{China}}}

\affil[9]{\orgname{Zhangjiang Institute for Advanced Study, Shanghai Jiao Tong University}, \orgaddress{\city{Shanghai}, \postcode{201210}, \state{Shanghai}, \country{China}}}


\abstract{
In quantum materials, nonlinear optical responses are highly sensitive to electronic structure and many-body interactions. Probing and manipulating such nonlinear processes is a complex and subtle endeavor, yet it offers deep insights into emerging physics and functionalities. Here, we report an anomalous ultrafast enhancement of second harmonic generation (SHG) in a strongly correlated cobaltite YbBaCo$_4$O$_7$. Above-bandgap femtosecond pumping increases SHG intensity by up to 60~\% within 200~fs, with the enhancement persisting for tens of picoseconds. The enhancement is strongly anisotropic, with substantial amplification of in-plane susceptibility tensors, whereas the out-of-plane counterpart shows negligible change. We attribute these anomalies to ultrafast photodoping-induced modulation of the on-site Coulomb repulsion, which dynamically renormalizes the band structure and selectively amplifies specific $\chi^{(2)}$ components. These nonlinear dynamics encode rich information about the orbital symmetries and energies of the states involved, opening new avenues for all-optical probing of electronic structure in strongly correlated materials.}

\maketitle

\section*{Introduction}
Second harmonic generation (SHG) and related nonlinear optical interactions are of critical importance in modern optics, serving as key mechanisms for generation, amplification, and frequency conversion of optical signals. Nonlinear responses are intrinsically determined by a material's electronic structure and symmetry~\cite{book1,ehhg1}. Recent advances have demonstrated ultrafast optical engineering as a powerful approach for manipulating nonlinear optical properties in quantum materials, offering new opportunities for photonic applications ~\cite{N2013,NP2018,N2019,N2021}. Beyond these, such nonlinear dynamics also hold great potential as a sensitive probe for symmetry breaking, collective excitations, and higher-order interactions, enabling the identification of hidden orders, photoinduced phase transitions, and Floquet phenomena~\cite{S2019,N2021,Torre2021,Zong2023}. 

However, photoexcitations in conventional noncentrosymmetric systems mainly cause processes like entropy increase, order parameter suppression and dipole moment screening, which are prone to suppress SHG response, as widely observed in various materials ~\cite{prl1991,pr1968,prx2019,scre2025}. Alternative routes include coherent Floquet engineering of electronic energy levels ~\cite{N2021}, mode-selective nonlinear phonon couplings~\cite{prl2017}, but normally with SHG suppression rather than amplification observed. Exploring routes to achieve ultrafast control and enhancement of nonlinear responses, particularly SHG, holds great promise for uncovering novel functionalities and underlying physics.

Strongly correlated systems provide a fertile platform for emerging nonlinear phenomena~\cite{NP2018,HHG-st-1,Zong2023}. In these systems, strong many-body interactions break the conventional quadratic energy-momentum relationship, leading to complex band dispersions that activate distinct nonlinear responses~\cite{CRO2017,Zhao2017,SC2018,prl22-127401,NJ2024,nlphonon1}. It has been suggested that Hubbard $U$, the on-site Coulomb repulsion central to strongly correlated systems, can be effectively modulated on femtosecond timescales using intense laser pulses~\cite{prlNi2018,NP2018,prb2015,GW,prbNi2020}. Such ultrafast modulation induces significant electronic band renormalization, including Hubbard band shifts and gap reduction, which in turn strongly modifies the nonlinear optical properties. However, experimental detection of such renormalization dynamics remains nontrivial and challenging. Previous investigations have predominantly relied on time-resolved angle-resolved photoemission spectroscopy~\cite{VOscr,TS2-doublon} and X-ray absorption spectroscopy~\cite{PhysRevX2022}, each with inherent limitations. Nonlinear spectroscopy offers a promising all-optical route that is convenient and bulk-sensitive~\cite{allopt,Zong2023}, yet its application remains largely underdeveloped.

Here, we demonstrate that ultrafast photodoping significantly enhances the SHG response of the correlated cobaltite YbBaCo$_4$O$_7$ (Yb114)~\cite{Yb114,CBCO1}. Comprehensive time-resolved and rotational anisotropy measurements reveal that near-resonant SHG components are substantially amplified, while off-resonant components show only weak changes. Such selective amplification reflects the orbital-specific dynamics, highlighting nonlinear spectroscopy as a probe of many-body correlations and orbital degrees of freedom. Two key features make Yb114 particularly suitable in this investigation. First, the non-centrosymmetric structure of Yb114 provides a substantial SHG response, which we use to probe photo-induced dynamics. Second, the UHB of Co-$3d$ orbitals constitutes the unoccupied states near the Fermi level~\cite{Y2,Y1,YO}. The interplay among $3d$ electrons, geometric frustration, and structural distortion gives rise to rich electronic properties in the system~\cite{Yb114,YBCO1,Ca1142,CBCO1}. Strong correlations allow photoexcitation to perturb the $3d$ electrons, altering the electronic structure and hence the nonlinear optical responses.

\section*{Static SHG of Yb114}
\begin{figure}[h]
	\centering
	\includegraphics[width=0.8\textwidth]{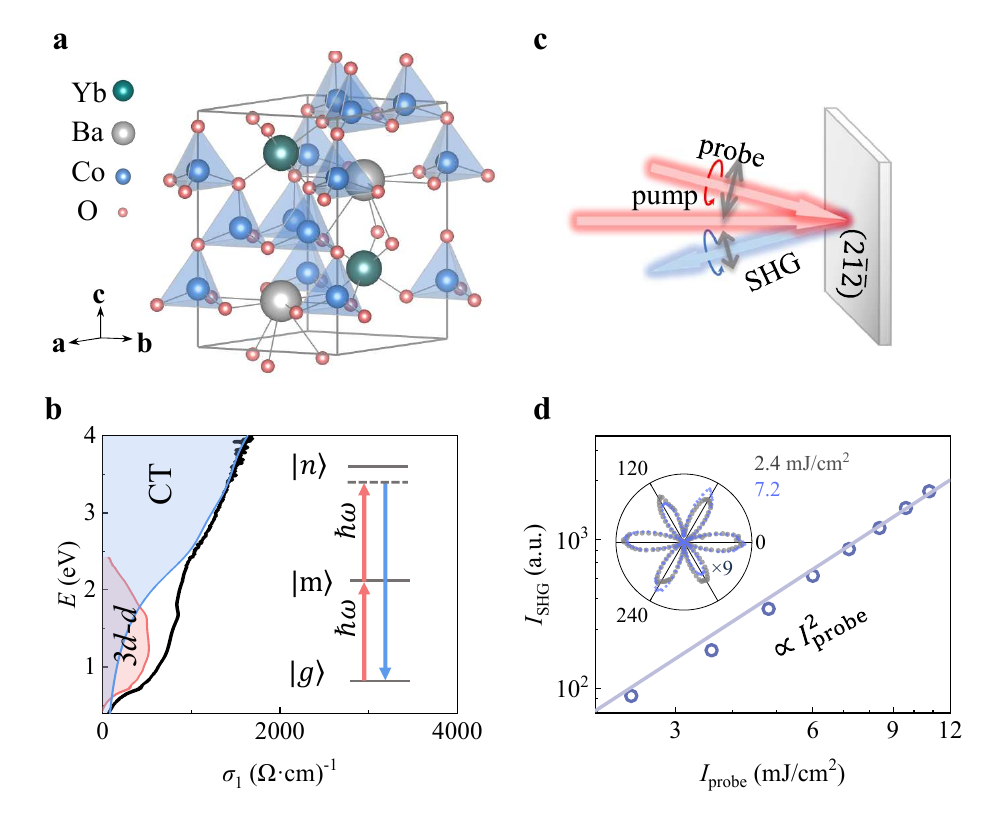} 
    \caption{\textbf{Static SHG of Yb114.} \textbf{a,} The crystal structure and \textbf{b,} Optical conductivity (black curve) with shaded areas indicating optical transitions. $3d$-$d$ (low-energy excitations) and CT (O $2p$-to-UHB charge transfer). Insert: black horizontal lines denote the initial, intermediate and final multi-electron states involved in the ED SHG process (red and blue arrows). \textbf{c,} The optical experimental setup for the SHG measurement. \textbf{d,} The static SHG intensity as the function of probe fluence. The solids is fitted line. Insert: The angular dependent SHG pattern (PP configuration) under different probe fluence at 800~$\mathrm{nm}$ without pump incident. Here, the SHG intensity for the 2.4~$\mathrm{mJ/cm^2}$ probe is magnified 9-fold to facilitate a clear comparison with the 7.2~$\mathrm{mJ/cm^2}$ curve.
	\label{fig:1} }
\end{figure}

The crystal structure of Yb114 features alternating Kagomé and triangular layers formed by CoO$_{4}$ tetrahedra (Fig.~\ref{fig:1}a). At room temperature, the compound adopts an asymmetric trigonal structure with space group $P31c$. The spontaneous polar distortions arise from displacements of Co ions within the CoO$_{4}$ tetrahedra, generating a net electric polarization along the crystallographic $c$-axis (point group $C_{3V}$). This polarization gives rise to an intrinsic second-order optical nonlinearity in the electric-dipole (ED) channel.

Fig.~\ref{fig:1}b presents the optical conductivity of Yb114 (See Methods). The spectrum reveals an optical gap onset of approximately 0.6~eV. The broad spectral weight centered near 1.2~eV covers the excitations from the Co $3d$ to UHB states~\cite{Y2,Y1,YO}. At higher energies ($>$~2~eV), another broad transition corresponds to O $2p$-to-UHB charge transfer (CT). For the SHG process (Fig.~\ref{fig:1}b insert) with a fundamental wavelength of 800~nm (1.55~eV), the resulting SHG wavelength of 400~nm (3.10~eV) falls near the center of the CT transitions.

SHG measurements were carried out on the naturally grown (2\,$\bar{1}$\,$\bar{2}$) plane of Yb114 sample under near-normal incidence in reflectance geometry, as illustrated in Fig.~\ref{fig:1}c. Rotational anisotropy second harmonic generation (RA-SHG) was performed in the \(P_{\text{in}}-P_{\text{out}}\) (PP) and \(P_{\text{in}}-S_{\text{out}}\) (PS) configurations. To avoid detecting direct SHG signals from the pump laser itself, a non-collinear arrangement was employed for the pump and probe beams. The measured static angular-dependent SHG patterns obey the canonical quadratic scaling as a function of the incident probe power (Fig.~\ref{fig:1}d).

For Yb114, the SHG response is dominated by an ED process of the form \( P_i(2\omega) = \chi_{ijk}^{(2)} E_j(\omega) E_k(\omega) \), where the second-order susceptibility tensor \( \chi_{ijk}^{(2)} \) governs the relationship between the incident (probe) electric field \( E_j(\omega) \) at frequency \( \omega \) and the induced polarization \( P_i(2\omega) \) at twice the incident frequency. The tensor \( \chi_{ijk}^{(2)} \) is intrinsically determined by the symmetry and the electronic structure of the system. Consequently, perturbation of the electronic state by optical pumping is expected to leave distinct signatures in the SHG response and hence in its susceptibility tensors.

\section*{Time-resolved SHG measurements}
\begin{figure}[h]
\centering
\includegraphics[clip, width=1\textwidth]{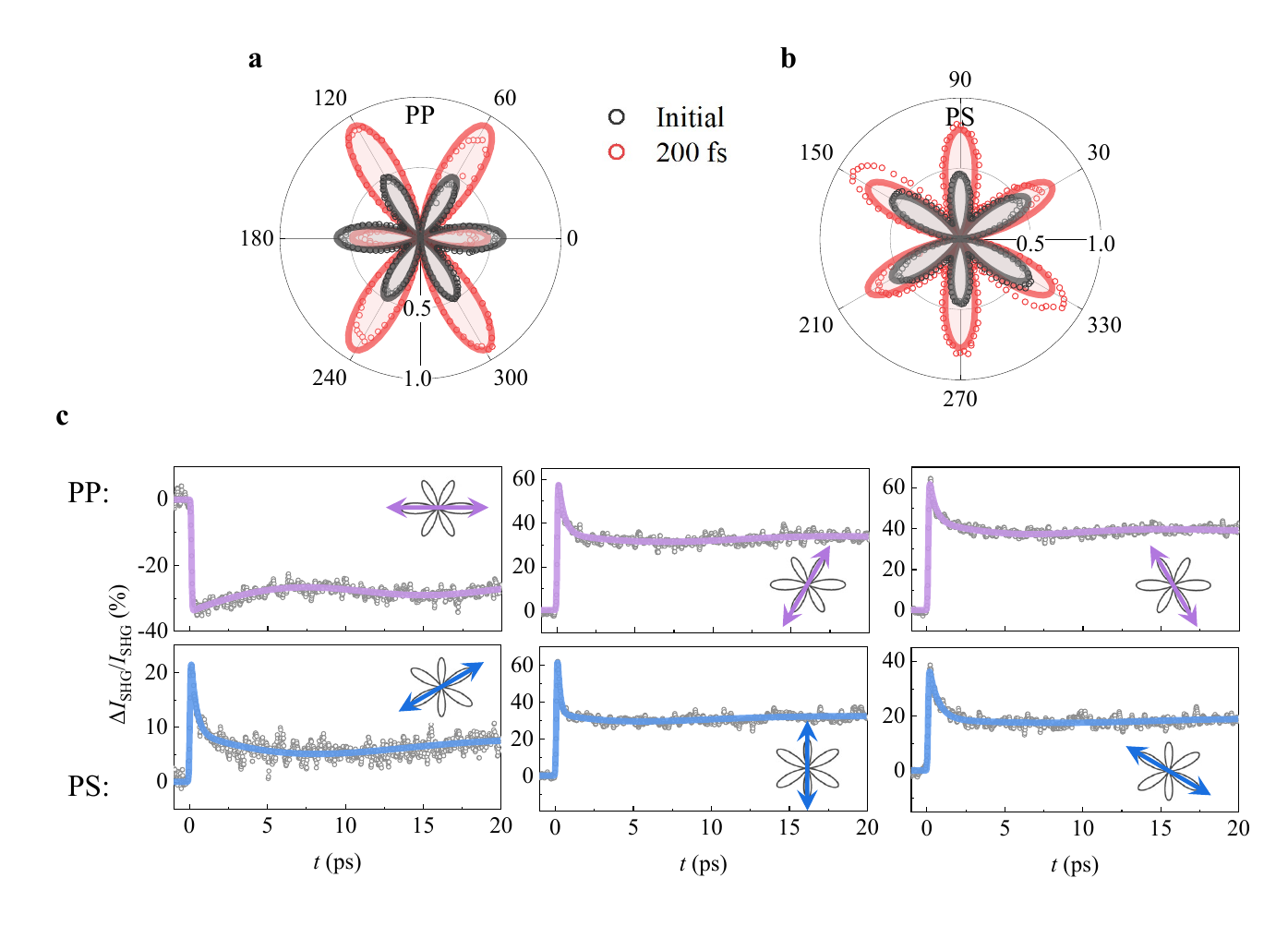}\\[1pt] 
\caption{\textbf{The rotational anisotropy and the dynamics of SHG signals under 800~nm pumping.} The initial state and excitation state SHG pattern for \textbf{a,} PP and \textbf{b,} PS configurations, respectively; \textbf{c,} The time resolved SHG in typical directions for PP (upper panels) and PS (bottom panels) modes respectively. Here, the pumping fluence is 10~$\mathrm{mJ/cm^2}$ and the incident probe wavelength is fixed at 800~nm. The solids are fitted lines.
\label{fig:2}}
\end{figure}

We first employed 800~nm pulses (1.55~eV, 50~fs) as the pump beam to investigate the SHG dynamics. This pump energy covers excitations from LHB to UHB, generating photodoped carriers. As a result, time-resolved rotational anisotropy SHG (tr-RA-SHG) measurements reveal an anisotropic SHG enhancement along different crystallographic directions. Fig.~\ref{fig:2}a and~\ref{fig:2}b show the RA-SHG signals collected at a fixed time delay of 200~fs. Most SHG lobes exhibit an intensity increase of 20--60~\%, whereas the lobes at $0^\circ$ and $180^\circ$ exhibit a suppression of approximately 30~\%.

Fig.~\ref{fig:2}c displays the tr-SHG signal as a function of pump-probe delay for representative probe polarizations corresponding to Fig.~\ref{fig:2}a and~\ref{fig:2}b. Upon pumping, the SHG response is rapid within ~200~fs. The ultrafast excitation process  indicates a pure electronic origin, which is further corroborated by the similarity between the tr-SHG and reflectivity change (Fig.~S3) It then undergoes a fast decay with a time constant of ~350~fs, and eventually evolves into a long-lived plateau persisting for tens of picoseconds (Fig.~S4). This anisotropic photoinduced SHG enhancement is not unique to Yb114, its isostructural counterpart YBaCo$_4$O$_7$, exhibits a comparable ultrafast SHG response under identical photoexcitation conditions (Fig.~S5).

In conventional noncentrosymmetric systems dominated by independent electrons, photoexcitation typically induces entropy increase, order-parameter suppression, and dipole screening, all of which tend to suppress SHG signals~\cite{prl1991,pr1968,scre2025}. The ultrafast SHG enhancement observed in Fig.2 is therefore markedly different from this conventional behavior. Importantly, such the pump-induced enhancement is fundamentally distinct from a trivial increase in SHG intensity caused solely by raising the probe fluence. Without pump excitation, the SHG pattern of Yb114 exhibits isotropic and proportional variations with increasing probe fluence (Fig.~\ref{fig:1}d). By contrast, the pump-induced enhancement reported here shows a pronounced anisotropic modulation.

\begin{figure}[h]
\centering
\includegraphics[clip, width=1\textwidth]{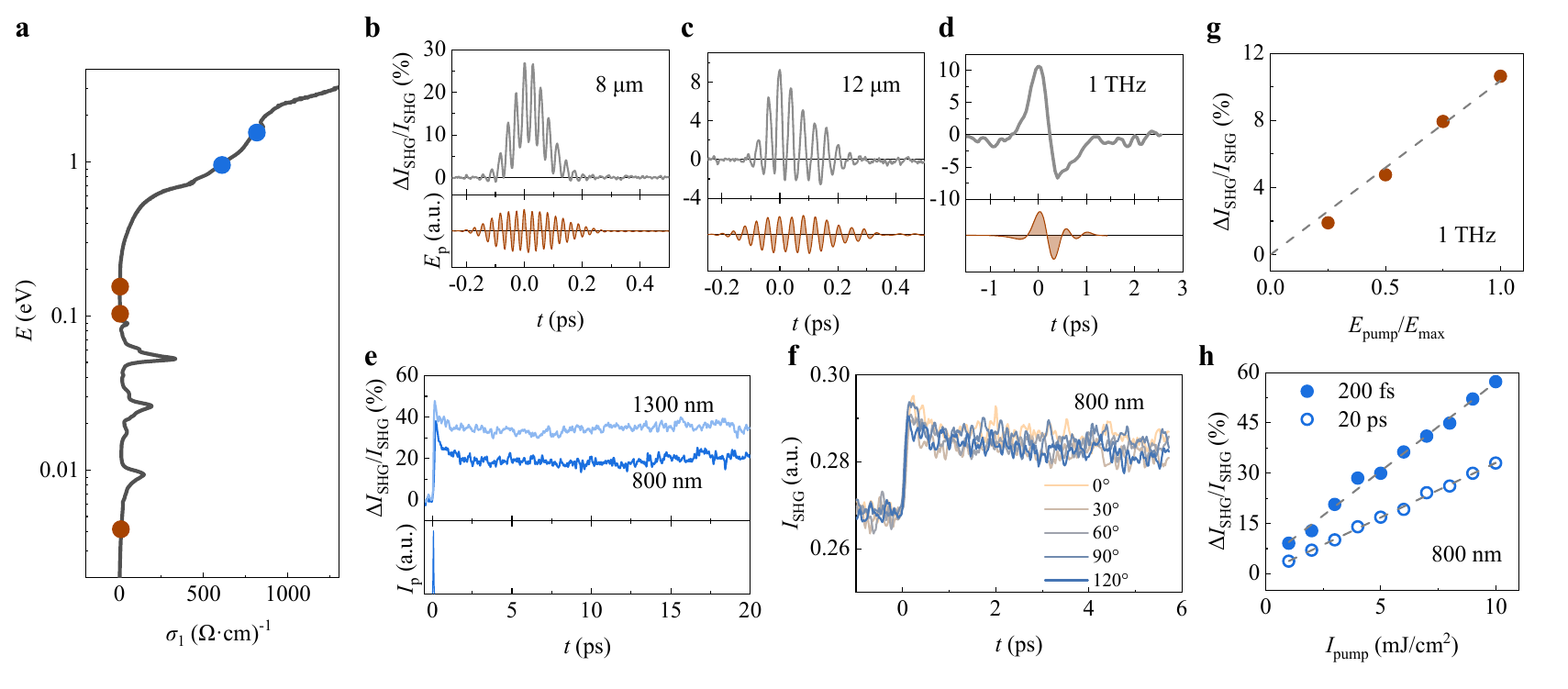}\\[1pt] 
\caption{\textbf{The pump wavelength dependence of SHG dynamics in Yb114.} \textbf{a,} Optical conductivity (black curve) and pumping lasers energy used (brown and blue dot). \textbf{b-e,} tr-SHG along 60$^\circ$ for pp configuration under various pumping incidents at \textbf{b,} 8~$\upmu$m, 1.3~mJ/cm$^2$, \textbf{c,} 12~$\upmu$m, 0.9~mJ/cm$^2$ , \textbf{d,} 1~THz, 1.1~MV/cm, \textbf{e,} NIR(800~nm, 6~mJ/cm$^2$ and 1300~nm, 6~mJ/cm$^2$). The brown line in \textbf{b-d,} depicts the electric field waveform of the THz/Mir pump pulse, the blue line in \textbf{e,} low panel depicts the intensity profile of the NIR pump pulse. The pump field and intensity are plotted in arbitrary units in the time domain; \textbf{f,} The pump polarization dependence of 800~nm SHG dynamics in Yb114 along PP-60$^{\circ}$. \textbf{g,} The field dependence of SHG under THz pump at $t$=0~ps; \textbf{h,} The fluence dependence SHG under 800~nm pump at 200~fs and 20~ps.
\label{fig:3}}
\end{figure}

To identify the necessary condition for SHG enhancement, we systematically investigated the pump-wavelength dependence by tuning the pump energy relative to the optical gap (Fig.~\ref{fig:3}a). Sub-gap pump excitation driven by mid-infrared (8~$\upmu$m, 12~$\upmu$m) and terahertz (1~THz) pulses yields exclusively transient SHG modulation that strictly follows the instantaneous pump electric field waveform and is confined within the pump pulse duration (Fig.~\ref{fig:3}b-d), ie., completely lacks a long-lived dynamical component. In sharp contrast, above-gap pumping at both 800~nm (1.55~eV) and 1300~nm (0.95~eV) produces sustained SHG enhancement that persists for tens of picoseconds longer than the pump pulse duration (Fig.~\ref{fig:3}e), with the 1300~nm excitation exhibiting the same anisotropic SHG dynamics as the 800~nm case (Fig.~S2).

Another notable difference lies in the response of SHG to the pump polarization. The sub-gap-induced modulation is highly pump-polarization-sensitive (SI-IV). This polarization dependence is consistent with an optical field effect, where the SHG modulation primarily arises from pump light-field coupling (Fig.~S8,~S9). By contrast, for above-gap-induced modulation, the SHG response is independent of pump polarization. As illustrated in Fig.~\ref{fig:3}f (and Fig.~S11), given a constant above-gap pump intensity, no measurable variation in SHG enhancement is observed under different pump polarizations.

Furthermore, sub-gap excitation, e.g., THz-driven modulation, induces SHG variations that scale linearly with the incident THz electric field intensity (Fig.~\ref{fig:3}g). Conversely, for above-gap pumping, the SHG enhancement scales linearly with pump fluence in the measured energy range. Fig.~\ref{fig:3}h presents the pump fluence dependence of SHG enhancement measured at two typical delay times of 200~fs and 20~ps. This linear fluence-dependent behavior differs fundamentally from that of sub-gap excitation.

Combined, these observations present the unique influence of above-gap pumping on the SHG response. These observed SHG features deviate from previously reported nonlinear modulation mechanisms, such as Floquet engineering and pump-polarization-dependent effects, including optical rectification and light-field-driven effects, among others (see Supplementary V for more details). Another potential factor to consider is the pump-induced change in the dielectric constant, which is primarily associated with the first-order linear response. However, the measured reflectivity change  $\Delta R/R$ (Fig.~S3) is found to be at least one order of magnitude smaller than the SHG variation. This change in reflectivity alone is far from sufficient to account for the significant increase in SHG intensity (see Supplementary III.2).

\section*{Discussion}
A few previous studies in correlated systems have demonstrated that nonlinear optical responses are highly sensitive to external conditions, albeit primarily under quasi-static conditions. For example, in the Mott insulator Ca$_2$RuO$_4$, researchers found that the HHG intensity in a specific spectral range exhibits a distinct scaling behavior, which is intrinsically linked to the interplay of electronic correlations and band renormalization dynamics~\cite{prl22-127401,cro2}. Whereas in the superconductor YBa$_2$Cu$_3$O$_7$, the electronic structure varies with temperature, leading to distinct nonlinear responses across different phase regimes~\cite{PNAS}.

In contrast, our current work focuses on a highly non-equilibrium process. It was predicted that photodoping can have a different dynamical effect on the electronic structure than quasi-static tuning~\cite{prb2015, GW}, and such ultrafast band structure modulation would affect the nonlinear optical property profoundly~\cite{prlNi2018,NP2018}. In our experiments, photoexcitation at 0.95~eV and 1.55~eV promotes electrons across the optical gap. Given that the SHG response involves the excitation and relaxation of UHB states, we consider the potential impact of photodoping effects on Yb114.

To establish the sensitivity of SHG to photodoping, we employ the band renormalization scenario for strongly correlated systems~\cite{prb2015, GW}, and consider the second-order nonlinear susceptibility within the sum-over-states framework:
\begin{equation}
\chi^{(2)}_{ijk}(2\omega;\omega,\omega) \propto \sum_{n,m} \frac{\langle g | \mu_{i}| n \rangle \langle n | \mu_{j} | m \rangle \langle m | \mu_{k} | g \rangle}{(E_{ng} - 2\hbar\omega - i\Gamma_n)(E_{mg} - \hbar\omega - i\Gamma_m)},
\label{eq:chi2}
\end{equation}
where $\mu$ denotes the electric dipole moment operator, $|g\rangle$ is the ground state, $|n\rangle,|m\rangle$ are final and intermediate states respectively, $E_{ng}$ and $E_{mg}$ are the corresponding resonant excitation energies, and $\Gamma$ are damping terms. The dynamical modulation of $U$ by photodoping impacts the SHG response mainly through band renormalization effects. As illustrated in Fig.~\ref{fig:4}a and discussed in Supplementary Note VII, the screening-induced redshift of the UHB significantly affects near-resonant SHG processes and could  enhance them.

To examine this, we extracted the independent tensor components from our experimental RA-SHG patterns. The resulting distinct responses of the SHG tensors  provide crucial insight into the microscopic origin of the photoexcitation dynamics. As shown in Fig.~\ref{fig:4}b, the relative changes of $\chi_{zxx}^{(2)}$ and $\chi_{xxz}^{(2)}$ are nearly identical and significantly larger than that of $\chi_{xxx}^{(2)}$, while $\chi_{zzz}^{(2)}$ remains almost unchanged. Fig.~\ref{fig:4}c further reveals that the induced anisotropy persists during  relaxation. The ratio $|\chi_{xxz}^{(2)}/\chi_{zzz}^{(2)}|$ increases by approximately 40~\% within 200~fs and remains partially elevated up to 20~ps. 

This anisotropic response reveals the orbital character of the involved states, as captured by group-theoretical selection rules~\cite{bookgroup,IC1998group}. In Yb114, the Co ions are known to adopt a high-spin configuration~\cite{Yb114,Y2,Y1,YO}. Considering the tetrahedral crystal field of the CoO$_4$ units, the Co $3d$ orbitals split into lower-energy $e$ ($d_{x^2-y^2}$, $d_{z^2}$-like) and higher-energy $t_2$ ($d_{xz}$, $d_{yz}$, $d_{xy}$-like) manifolds, with the $e$ orbitals being mostly occupied. The UHB is primarily composed of unoccupied $t_2$ orbitals, with a partial contribution from the $d_{x^2-y^2}$ orbitals. 

In the $C_{3v}$ point group, $\chi_{zzz}^{(2)}$ transforms as the $A_1$ irreducible representation~\cite{bookgroup,IC1998group}. This component requires significant $c$-axis orbital character (i.e., $d_{z^2}$) for its $\mu_z$ transition dipoles. Since the UHB predominantly consists of in-plane orbitals, its renormalization has little effect on $d_{z^2}$-like orbitals and thus on $\chi_{zzz}^{(2)}$ (see Supplementary Note VI for more details). 

By contrast, $\chi_{zxx}^{(2)}$ and $\chi_{xxz}^{(2)}$ share the same two-photon resonance denominator involving the degenerate $d_{xz}$ and $d_{yz}$. Under near-resonant conditions, where the SHG photon energy at 3.1~eV lies close to the CT transition centered at approximately 3.3~eV (Fig.~S12), pump-induced changes in these orbitals can significantly affect the corresponding SHG tensors. Specifically, upon above-gap pumping, photodoping screens the on-site Coulomb repulsion $U$, inducing a downward redshift of the UHB. This brings the $d_{xz}$ and $d_{yz}$ states closer to the $2\omega$ resonance condition, selectively amplifying $\chi_{zxx}^{(2)}$ and $\chi_{xxz}^{(2)}$. Meanwhile, the $d_{x^2-y^2}$ orbital, which is a primary contributor to $\chi_{xxx}^{(2)}$, remains off-resonance (near the bandgap edge) from the CT center. Consequently, the pump-induced change in $\chi_{xxx}^{(2)}$ is substantially smaller than that of the near-resonant components .

Within the band-renormalization framework~\cite{prb2015, GW}, the SHG enhancement can be further quantitatively understood. The photodoping concentration $\delta$ reduces the effective Hubbard interaction as $U_{\text{eff}}(\delta) = U_0 - \beta\delta$, and the charge gap contracts accordingly $E_g(\delta) = E_g(0) - \alpha\delta$. Since $\delta \propto F$ (pump fluence), the resonance energy for a dominant channel near $2\hbar\omega$ shifts as $E_{\text{res}}(F) = E_{\text{res}}(0) - \tilde{\alpha}_{\text{res}} F$. The second-order susceptibility then takes the resonant form $\chi^{(2)}(2\omega) \propto \frac{B}{E_{\text{res}}(F) - 2\hbar\omega - i\gamma}$, with $B$ containing matrix elements and $\gamma$ a damping term. This reduces the detuning $\Delta_{\text{res}} = E_{\text{res}}(0) - 2\hbar\omega$, bringing the system closer to resonance and yielding the observed SHG enhancement. This interpretation is corroborated by numerical simulations (Fig.~\ref{fig:4}b). For $\chi^{(2)}_{xxz}$ and $\chi^{(2)}_{zxx}$, given a near-resonant condition with $\Delta_{\text{res}} = 0.2$~eV and $\gamma = 0.15$~eV, the estimated redshift of $E_{\text{res}}$ is 0.17~eV at a fluence of 10~mJ/cm$^{2}$, which yields an approximately 63~\% enhancement. 

The above resonance scenario accounts for the instantaneous SHG enhancement amplitude, while the long-lived response points to an additional stabilization mechanism. The SHG response persists for tens of picoseconds (Fig.~\ref{fig:4}c), far longer than typical electron decay times, suggesting coupling to a slower degree of freedom. Considering the strongly polar CoO$_4$ tetrahedra in Yb114, photocarriers are expected to become self-trapped, stabilizing the renormalized band structure~\cite{polaron}. This assignment is further supported by the absence of long-lived enhancement under sub-gap excitation, where real carriers are not generated.

\section*{Conclusion and Outlook}
Nonlinear optical properties in correlated systems are exquisitely sensitive to external perturbations. We exploit this sensitivity in the noncentrosymmetric cobaltite Yb114. Ultrafast photodoping drives a pronounced, anisotropic enhancement of SHG, reaching up to 60~\%. This contrasts starkly with the SHG suppression typically observed in conventional systems. The enhancement is highly selective, with the SHG components under near-resonant conditions substantially amplified, while the off-resonant component remains essentially unchanged. These tensor dynamics encode rich information about the orbital symmetries and energies involved.

Our findings not only demonstrate the distinctive nonlinear optical responses in strongly correlated systems, but also establish the nonlinear dynamics as a sensitive probe for resolving orbital states. Further developments in broadband coherent spectroscopy and multi-order nonlinear detection should advance this research in quantum materials science. Given the rich quantum phenomena hosted by strongly correlated systems, this nonlinear approach opens new opportunities for probing magnetic, electronic, and superconducting states through ultrafast correlation dynamics.

\clearpage

\clearpage

\clearpage

\clearpage
\begin{figure}[h]
\centering
\includegraphics[clip, width=0.8\textwidth]{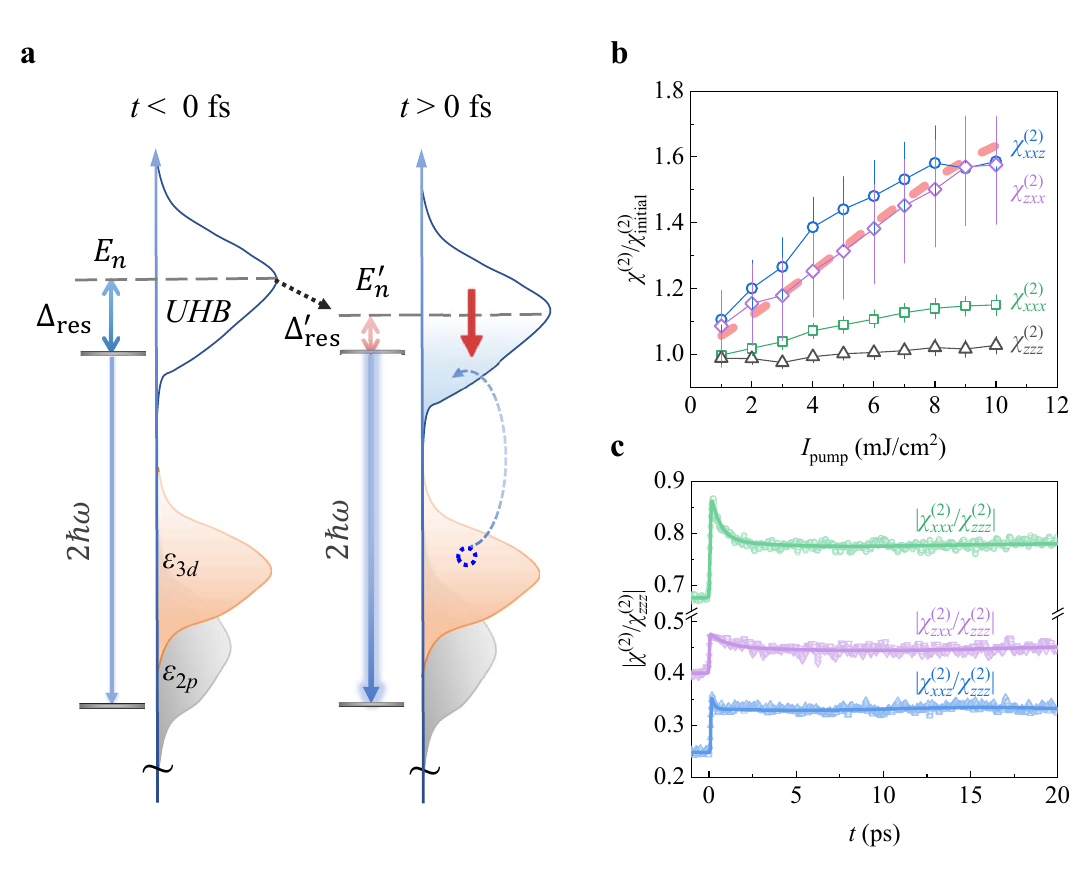}\\[1pt] 
\caption{\textbf{Orbital-selective enhancement of second-order susceptibility.}\textbf{a,} Schematic of photoinduced band renormalization. Left pannel: before pumping. Right pannel: after above-gap pumping, the UHB shifts downward (redshift) towards resonance with the SHG photon energy. \textbf{b,} Pump-fluence dependence of normalized $\chi^{(2)}$ components at 200~fs. $\chi_{xxz}^{(2)}$ increases linearly up to $\sim$60~\%, while $\chi_{zzz}^{(2)}$ shows negligible change. The red dashed line represents the simulation result. \textbf{c,} Typical time evolution of the tensor ratio $|\chi^{(2)} / \chi_{zzz}^{(2)}|$. The ratio rises within 200~fs and sustains a long-lived plateau, demonstrating persistent anisotropy in the photoinduced correlation renormalization. Error bars represent standard deviations from multiple RA-SHG pattern fittings. 
\label{fig:4}}
\end{figure} 

\clearpage

\section*{Methodes}
\subsection*{Sample growth}
Single crystals of YbBaCo$_4$O$_7$ were grown using a BaO-CoO self-flux system. BaCO$_3$ and Co$_3$O$_4$ were mixed in a 2:1 molar ratio to reach the eutectic composition of the flux, with Yb$_2$O$_3$ added at 7.5--12.5~wt\%. The mixture was placed into Al$_{2}$O$_{3}$ crucibles, calcinated at 1000~$\mathrm{^{\circ}C}$ for 12~hours, heated to 1300~$\mathrm{^{\circ}C}$ and held for 3~hours, finally cooled to 1050~$\mathrm{^{\circ}C}$ at a rate 1~$\mathrm{K/h}$. The obtained single crystals were hexagonal bulk crystals of millimeter size with shiny surfaces. The crystal structure was confirmed by single-crystal X-ray diffraction.

\subsection*{Laser and THz generation}
The 800~nm and THz excitation setup employed a Spitfire Ace Advanced Ti:Sapphire regenerative amplifier (Spectra-Physics). The output laser delivered 800~nm linearly polarized light at a repetition rate of 1~kHz. The THz pulses were generated in an MgO-doped LiNbO$_3$ crystal using wavefront tilt technology.

The 1300~nm and MIR excitation setup used a PHAROS-SP femtosecond laser (Light Conversion). An ORPHEUS collinear optical parametric amplifier produced either 800~nm and 1300~nm or 800~nm and MIR linearly polarized light at a repetition rate of 50~kHz. For 1300~nm interband excitation, the repetition rate was reduced to 5~kHz.

The pump light waveforms were obtained using electro-optic sampling. The THz field signals were calibrated with a GaP crystal, and the MIR field signals were calibrated with a GaSe crystal. The near-infrared pulse width was measured using an autocorrelator.

\subsection*{tr-RA SHG and linear experiments}
The optical measurements were performed in a nearly normal incidence geometry for both reflectivity and SHG. The reflected light signal was collected using a photodetector coupled to a lock-in amplifier. A chopper was placed in the pump beam to measure the reflectivity change ($\Delta R$). The positive and negative signal relations were calibrated using a single-crystal silicon standard sample. The static reflectivity $R$ was obtained from the oscilloscope.

The SHG signal was collected using a photomultiplier tube (PMT) coupled to a lock-in amplifier. Rotational anisotropy SHG (RA-SHG) measurements were performed by synchronously rotating the polarization of the incident and reflected polarizers. The reflected polarizer was set to parallel for the P$_{in}$-P$_{out}$ (PP) configuration and perpendicular for the P$_{in}$-S$_{out}$ (PS) configuration. To measure the absolute SHG intensity $I_{\mathrm{SHG}}$ rather than its relative variation, a chopper was placed in the probe beam path and referenced to the lock-in amplifier.

For Optical spectral properties, near-normal incidence reflectance spectra were collected using a Bruker 80v/S Fourier-transform infrared spectrometer in the 50--32,000~cm$^{-1}$ range. To minimize surface scattering effects, an in situ overcoating technique was employed during the reflectance measurements. The optical constants were subsequently extracted from the reflectivity spectra using Kramers--Kronig transformation and the Fresnel formula for a semi-infinite medium.

\backmatter

\bmhead{Supplementary information}

\bmhead{Acknowledgements}
This work was supported by National Natural Science Foundation of China (Grant No.12274033, 12488201, 12404166, 12574349), National Key Research and Development Program of China (2022YFA1403901, 2024YFA1408700), and the Synergetic Extreme Condition User Facility (SECUF, https://cstr.cn/31123.02.SECUF).

\section*{Declarations}
The authors declare no competing interests.


\bibliography{sn-bibliography}

\end{document}